\documentstyle[twocolumn,pra,aps,epsf]{revtex}
\def\epsfbox#1{} 
\begin{document}
             \title{ Analysis of the Stern-Gerlach Measurement
\thanks{SPS-JNU preprint}}

            \author{A. Venugopalan, Deepak Kumar and R. Ghosh}
                    \address{School of Physical Sciences,
                        Jawaharlal Nehru University,
                         New Delhi - 110067, INDIA}
\draft
\maketitle
\begin{abstract}
     A  dynamical  model  for the collapse of the wave function in a quantum
measurement  process is proposed by considering the interaction of a quantum
system  (spin-1/2)  with a macroscopic quantum apparatus interacting with an
environment  in  a  dissipative manner. The dissipative interaction leads to
decoherence  in  the  superposition  states  of  the  apparatus,  making its
behaviour  classical  in  the sense that the density matrix becomes diagonal
with  time.  Since  the  apparatus  is also interacting with the system, the
probabilities  of  the  diagonal  density matrix are determined by the state
vector  of the system. We consider a Stern-Gerlach type model, where a spin-
1/2  particle  is in an inhomogeneous magnetic field, the whole set up being
in contact with a large environment. Here we find that the density matrix of
the  combined  system and apparatus becomes diagonal and the momentum of the
particle  becomes correlated with a spin operator, selected by the choice of
the  system-apparatus interaction. This allows for a measurement of spin via
a  momentum  measurement  on  the  particle with associated probabilities in
accordance with quantum principles.
\end{abstract}
\begin{narrowtext}
\section{Introduction}

After about eighty years of remarkable success of quantum mechanics, the
understanding of its measurement aspects remains poor and debatable. The
root of the difficulty is that while the state vector of a quantum system
has a deterministic evolution, the measurable properties of the system can
only be predicted probabilistically.  The linearity of quantum equations
allows for solutions which are a linear superposition of some basis states,
each of which may correspond to one of the different values of a dynamical
variable.  So the definitive outcome that is obtained in a given
measurement of this variable can only be accounted for by the notion of the
'collapse' of the wavefunction to one of the basis states contained in the
wavefunction.  Though the question as to which state the wavefunction
collapses to can be answered probabilistically, quantum mechanics contains
no mechanism for the collapse. Bohr \cite{1} and von Neumann \cite{2}
postulated that this process occurs when the quantum system comes into
contact with an apparatus which must be described classically. Moreover,
since the state vector can be expressed as a linear superposition of basis
states in any number of ways, the interaction between the classical
apparatus and the system decides which property would be measured.
According to Bohm \cite{3} this implies that the classical properties as we
observe them are contained only as potentialities in the state vector.
Many authors have expressed dissatisfaction with this dichotomy between the
quantum world and the classical world.  Are there really two kinds of
systems? If so, is there a way to describe the combined dynamics of quantum
and classical systems to exhibit the so called 'collapse'? No satisfactory
answer to these issues has emerged even after intensive efforts by a large
number of workers \cite{4,5,6,7,8,9}.

An interesting line of investigation to resolve this issue was initiated by
Zeh \cite{7}, who observed that the measurement apparatus, being always a
macroscopic object, has closely spaced energy levels which make it very
susceptible to the influence of the environment. The environment consists
of a large number of degrees of freedom and its interaction with the
apparatus causes decoherence in the quantum evolution of the latter.  This
decoherence is quite a general feature whenever one considers the
interaction between a small quantum system and a large one and monitors the
density matrix of the small system only \cite{10}. Here the off-diagonal
elements of this reduced density matrix decay in time. However such decays
seem dissipative only for small times as they are actually arising due to
the superposition of a large number of harmonic terms, viz.$\sum_{j=1}^{n}
a_{j} e^{i\omega_{j}t}$ , which, under the condition that $\omega_{j}$ are
closely spaced, give rise to apparent decays for time $t \ll T$, where $T$
is a recurrence time which can be astronomically large, even under mild
conditions of $n \sim 1000$ and $\Delta\omega\sim 10^{-5}$. Thus a quantum
system coupled to an environment consisting of a large number of degrees of
freedom behaves like a classical system in the sense that at time scales of
interest, its density matrix is driven diagonal. Since the diagonal density
matrix is interpretable in classical terms, several authors have examined
the effect of environmental interactions and state reduction in a variety
of systems and circumstances \cite{7,8,9,10,11,12,13,14,15,16,17,18}.

     In  this paper our goal is to further this line of inquiry by examining
the  process  of  measurement  in the following way. We consider a canonical
spin-1/2  quantum  system  interacting  with  an apparatus, which in turn is
interacting  with  an  environment. The basic idea of the scheme is to study
the following two essential features of the quantum measurement process. The
first  is the establishment of correlations between states of the system and
the  apparatus.  The  second  is  the reduction of the density matrix of the
macroscopic apparatus to a diagonal form dictated by the correlations of the
apparatus with the system. The second aspect is achieved here by interaction
of  the apparatus with the environment. Thus through this scheme one is able
to  study  the  interaction of a quantum system with a macroscopic apparatus
within  the pure realm of quantum mechanics and provide a realization of the
early ideas of Bohr.

We consider the set up of a Stern-Gerlach apparatus, i.e., a spin-1/2
particle in the presence of an inhomogeneous magnetic field. A measurement
of spin is made by studying the position or momentum of the particle. Such
a model has been considered in some detail by Bohm \cite{3}. Here spin
plays the role of the system, and the position/momentum degrees of freedom
of the particle that of the apparatus.  The particle is further coupled to
an environment via its position, and this coupling is intended to drive its
translational behaviour classical to perform a measurement. The effect of
the environment on the translational degrees of freedom is taken into
account via the density matrix equation which incorporates both the quantum
evolution and the stochastic Fokker-Planck type evolution arising due to
the environmental interaction.  This equation has been obtained in a
variety of ways in recent literature \cite{10,13,15}.  We solve this
equation exactly by including an appropriate system-apparatus interaction
and show that in this case the reduced density matrix is driven diagonal at
long times in the momentum space and has the desired correlation with a
component of spin.

     The  remaining  paper  is organized as follows. Section II contains the
calculations  and  analysis  of  the  density  matrix  for the Stern-Gerlach
apparatus,  and section III summarizes our various observations arising from
these calculations.

\section{Density matrix for the Stern-Gerlach apparatus}

     We   consider   a  Stern-Gerlach  type  set-up  for  investigating  the
measurement  of  spin.  The Hamiltonian of the combined system/apparatus and
environment is \cite{19}
\begin{equation}
H^{SAE} = p^{2}/2m + \lambda\sigma_{z} + \epsilon x
\sigma_{z}+H^{AE}+H^{E}.\label{1}
\end{equation}

Here $x$ and $p$ denote the position and momentum (taken in one dimension
for convenience) of the particle of mass $m$, $\lambda\sigma_z$ the
Hamiltonian of the system, $\epsilon$ the product of the field gradient and
the magnetic moment of the particle, $H^{AE}$ the interaction of the
environmental degrees of freedom with $x$, and $H^{E}$ denotes the
Hamiltonian for the environmental degrees of freedom. Since the problem of
motion of a spinless particle in simple potentials and in interaction with
the environment has been studied at great length in recent literature
\cite{10,13,14,15}, we draw upon this work to deal directly with a reduced
density matrix equation for the particle in which environmental degrees of
freedom have been traced over.  Though the density matrix equation has been
derived in a number of ways, the derivation which is in the spirit of the
present work was given by Caldeira and Leggett using Feynman-Vernon path
integral approach \cite{10}. In this method the path integral expression
for the density matrix is written for the Hamiltonian of (\ref{1}), and then
the degrees of the environment which consists of oscillators are integrated
out.  In the limit of high temperature (weak coupling) the expression for
the reduced density matrix is seen to be a solution of the above mentioned
density matrix equation. This  density matrix equation can be thought of as
a Markovian limit of a Generalized Master equation, and thus its validity
lies in the large time domain, $t \gg t_{c}$, where $t_{c}$ is a short
relaxation time associated with the environment. This limit is needed here
as the earlier
work \cite{15} shows that the solution then evolves to a classical
stochastic distribution. The tracing over environmental degrees of freedom
may be construed as if one is not dealing with one measurement process.
However, we feel that this is an essential characteristic of the
interaction of a system with a macroscopic object and may be viewed in the
same spirit as a single measurement is regarded as an ensemble average for
macroscopic systems.

We look at the time evolution of the density matrix in the $|s,x\rangle$
representation, where $|s\rangle$ refers to the eigenstates of $\sigma_{z}$
and $|x\rangle$ are the position states. Corresponding to the four elements
of the spin space $(\uparrow\uparrow, \downarrow\downarrow,
\uparrow\downarrow,
\downarrow\uparrow)$, the equations for the elements of the reduced
density matrix $\rho_{ss'}(x,y,t)$ for our Hamiltonian are:
\begin{eqnarray}
{\partial\rho_{ss'}(x,y,t)\over \partial t} &=& \biggl[ {-\hbar\over 2im}
\left({\partial^{2}\over\partial x^2} - {\partial^{2}\over\partial y^{2}}
\right) \nonumber\\
&&- \gamma(x-y) \left({\partial\over\partial x}
- {\partial\over\partial y}\right)- {D\over 4\hbar^2}(x-y)^{2} \nonumber\\
&&+ {i\epsilon(xs-ys')\over\hbar}
+ {i\lambda(s-s')\over\hbar}\biggr]\rho_{ss'}(x,y,t),\nonumber\\
\label{2}
\end{eqnarray}
where $s, s' =$ + 1 (for $\uparrow$) or -1 (for $\downarrow$). Here
$\gamma$ is the Langevin friction coefficient and $D$ has the usual
interpretation of the diffusion coefficient. In the case of a heat bath of
harmonic oscillators at temperature $T$, $D = 2\gamma mk_{B}T$. It is more
convenient to work with variables: $r \equiv (x-y)$, $R \equiv (x+y)/2$.
Then the spin diagonal density matrix $\rho_{d}$, and the spin off-diagonal
density matrix $\rho_{od}$ obey the equations:
\begin{equation}
{\partial\rho_{d}(R,r,t)\over\partial t} = {-\hbar\over im}{\partial^{2}
\rho_{d}\over\partial R\partial r} - \gamma r {\partial\rho_{d}\over
\partial r}-{Dr^{2}4\hbar^2}\rho_{d} \pm{i\epsilon r\over\hbar}
\rho_{d}, \label{3}
\end{equation}
where the `+' sign in the last term corresponds to $\rho_{\uparrow
\uparrow}$ and `-' to $\rho_{\downarrow\downarrow}$, and
\begin{eqnarray}
{\partial\rho_{od}(R,r,t)\over\partial t} &=& {-\hbar\over im}{\partial^{2}
\rho_{od}\over\partial R\partial r} - \gamma r {\partial\rho_{od}\over
\partial r}-{Dr^{2}4\hbar^2}\rho_{od} \nonumber\\
&&\pm{2i\epsilon r\over\hbar}
\rho_{od} \pm{2i\lambda r\over\hbar}\rho_{od} , \label{4}
\end{eqnarray}
where  the upper signs in the last two terms correspond to $\rho_{\uparrow
\downarrow}$  and the lower ones  to $\rho_{\downarrow\uparrow}$. To solve
these equations, it is convenient to take a partial
Fourier transform in the variable $R$:
\begin{equation}
\rho (Q,r,t)=\int^{\infty}_{-\infty}\exp(iQR)\rho (R,r,t) dR . \label{5}
\end{equation}
The  equations (\ref{3}) and (\ref{4}) simplify to a pair of first-order
partial
differential equations:
\begin{equation}
{\partial\rho_{d}(Q,r,t)\over\partial t} = \left({\hbar Q\over m}-\gamma r
\right) {\partial\rho_{d}\over\partial r}-{Dr^{2}\over 4\hbar^2}\rho_{d}
\pm{i\epsilon r\over\hbar}\rho_{d} , \label{6}
\end{equation}
\begin{eqnarray}
{\partial\rho_{od}(Q,r,t)\over\partial t} &=& \left({\hbar Q\over m}-\gamma r
\right) {\partial\rho_{od}\over\partial r}\nonumber\\
&&-{Dr^{2}\over 4\hbar^2}\rho_{od}
\pm{2\epsilon\over\hbar}{\partial\rho_{od}\over\partial Q}
 \pm{i\lambda r\over\hbar}\rho_{od} , \label{7}
\end{eqnarray}
Such equations, being of first order, can be solved by the method of
characteristics \cite{20}.  The physical significance of the solution can
be clearly understood if we choose the initial condition to be the
following Gaussian wave packet of width $\sigma$ and mean momentum
$\overline{p}$:
\begin{equation}
\psi (x,0) = {1\over(\sigma\sqrt \pi)^{1/2}} \exp(i\overline px
-x^{2}/2\sigma^{2}). \label{8}
\end{equation}
The solution for (\ref{7}), i.e., the spin {\sl off-diaginal} elements
of the density matrix, for the initial conditions of (\ref{8})
is (see appendix):
\begin{eqnarray}
\rho_{od}(Q,r,t)&=&\exp\left( {\tau^{3}\epsilon^{2}D\over 3m^{2}
\gamma^{5}\hbar^{2}}\right)\exp\left({\pm 2i\lambda
t\over\hbar}\right)\nonumber\\
&&\exp\biggl[-{1\over
4}\left({D\over\hbar^{2}\gamma} (1-e^{-2\tau})+{1\over\sigma^{2}}
e^{-2\tau}\right) r^{2}\nonumber\\
&&+\Bigl[ i\overline pe^{-\tau}\mp{\epsilon\tau
e^{-2\tau}\over\gamma^{2} m\sigma^{2}} \mp {\epsilon D\over
2\hbar^{2}m\gamma^{3}} (\tau(1-e^{-2\tau})\nonumber\\
&&-2(1-e^{-\tau}))\pm
{D\epsilon\tau \over\hbar^{2}m\gamma}\Bigr]r -\Bigl({1\over
4\sigma^{2}}(1-e^{-\tau})^{2}\nonumber\\
&&+{D\over 8\hbar^{2}\gamma}(4\tau-3+4e^{-\tau}-e^{-2\tau})\Bigr)
r^{2}_{Q}\nonumber\\
&& + \Bigl( ip(1-e^{-\tau})-{1\over 4\sigma^{2}}
(2r\pm{4\epsilon\tau\over \gamma^{2}m})(1-e^{-\tau}) \nonumber\\
&&-{Dr\over 4\hbar^{2}\gamma}(1-e^{-\tau})^{2} \pm  {\epsilon D\over
2\hbar^{2}m\gamma^{3}}(1-e^{-\tau})(\tau(1\nonumber\\
&&-e^{-\tau})-2)
\pm  {D\epsilon\tau{2}\over 2\hbar^{2}m\gamma^{2}} \mp
{D\epsilon\tau\over\hbar^{2}m\gamma^{2}}\Bigr) r_{Q}\biggr]\nonumber\\
&&\times\exp\left(-{Q^{2}\sigma^{2}\over 4}-
\left({\epsilon\tau\sigma\over\hbar\gamma}\right)^{2}\right),\label{9}
\end{eqnarray}
where $\tau = \gamma t$ , and
\begin{equation}
r_{Q} = Q\hbar /m\gamma\pm 2\epsilon\tau /m\gamma^{2}
\pm 2\epsilon /m\gamma^{2}.\label{10}
\end{equation}
The solution has a factor going as $e^{-A\tau^{3}}$ which drives the entire
expression to zero with time, independent of all other arguments in the
density matrix. This means that the density matrix is driven diagonal in
the spin-space. The time scale over which this happens is given by
\begin{equation}
\tau_{s} = \left({3m^{2}\gamma^{5}\hbar^{2}\over\epsilon^{2}D}\right)^{1/3}.
\label{11}
\end{equation}
The the solution of (\ref{6}), i.e, for the {\sl spin-diagonal}
is (see appendix for details)
\begin{eqnarray}
\rho_{d}(Q,r,t)&=&\exp\Bigl[ i\overline r_{Q}-r^{2}_{Q}/4\sigma^{2}
-Q^{2}\sigma^{2}/4 \nonumber\\
&&+ i\overline p(r-r_{Q}e^{-\tau}-{1\over 4}\sigma^{2} \{
(r-r_{Q})^{2}e^{-2\tau} + 2r_{Q}\nonumber\\
&&(r-r_{Q})e^{-\tau}\}
- D/(4\hbar^{6}\gamma) \{ r^{2}_{Q}\tau +2r_{Q}\nonumber\\
&&(r-r_{Q})(1-e^{-\tau})
+ (r-r_{Q})^{2}(1-e^{-2\tau})/2\} \nonumber\\
&&\mp i\epsilon /\hbar\gamma \{
r_{Q}\tau + (r-r_{Q})(1-e^{-\tau})\}\Bigr] , \label{12}
\end{eqnarray}
where $r_{Q} = \hbar Q/m\gamma$. To understand the
measurement aspect implied by this  solution,  we  consider  the  solution
in the momentum representation,
i.e.,
\begin{equation}
\rho_{d}(\overline u,\overline v,t) = \int\rho_{d}(x,y,t)
e^{i(\overline ux+\overline vy)} dx dy . \label{13}
\end{equation}
This  is obtained by taking a Fourier transform with respect to the variable
r in (\ref{12}) and identifying $Q = \overline u-\overline v$ and $q = (
\overline u + \overline v)/2$. This solution is
\begin{eqnarray}
\rho_{d}(Q,q,t)&=&2\sqrt{{\pi\over N(\tau)}}\exp\biggl[ {-1\over N(\tau)}
\Bigl[q + \overline pe^{-\tau}\nonumber\\
&&\mp{\epsilon\over\hbar\gamma}(1-e^{-\tau})
+{i\hbar Q\over 2\sigma^{2}m\gamma}e^{-\tau}(1-e^{-\tau})\nonumber\\
&&-{iQD\over 4\hbar\gamma^{2}m}(1-e^{-\tau})^{2}\Bigr]^{2}-
\Bigl[\left(\hbar\over 4\sigma m\gamma\right)^{2}\nonumber\\
&&(1-e^{-\tau})^{2}+{\sigma^{2}/4}
+ {D\over 2m^{2}\gamma^{3}}(2\tau -3+4e^{-\tau}\nonumber\\
&&-e^{-2\tau})\Bigr]Q^{2}
+\Bigl[{i\overline p\hbar\over m\gamma}(1-e^{-\tau})
\mp {i\epsilon\tau\over\ m\gamma^2}\nonumber\\
&&\pm{i\epsilon\over m\gamma^2}
(1-e^{-\tau})\Bigr]Q\biggr], \label{14}
\end{eqnarray}
where
\begin{equation}
N(\tau)\equiv (D/2\hbar^{2}\gamma)(1-e^{-2\tau})+(1/\sigma^{2})e^{-2\tau}.
\label{15}
\end{equation}
Now we see that the momentum off-diagonal components ($Q\neq 0$) vanish
with time, reducing $\rho_{d}$ to the diagonal form \cite{10,13,14,15}. The
time scale over which this happens is given by
\begin{equation}
\tau^{-1}_{d}={DQ^2\over m^{2}\gamma^3}. \label{16}
\end{equation}
The  momentum  distribution  function  can  be  obtained  by  looking at the
diagonal elements of the density matrix (\ref{14}), i.e., for $Q = 0$ and $q =
\overline u$ :
\begin{eqnarray}
\rho_{d}(0,\overline u,t) \equiv |\psi(\overline u)|^{2}&=&2
\sqrt{\pi\over N(\tau)}\exp\biggl({-1\over N(\tau)}\{\overline u +
\overline pe^{-\tau}\nonumber\\
&&\mp{\epsilon\over\hbar\gamma}(1-e^{-\tau})\}^2\biggr).
\label{17}
\end{eqnarray}
This has the classical Ornstein-Uhlenbeck form with the spin-dependent
drift caused by the field. In the large $t$ limit, the momentum
distribution is centered around $\epsilon /\hbar\gamma$ ($-\epsilon
/\hbar\gamma$) for up (down) spin. Thus we see that the measurement of
momentum of the particle can determine the spin. It is also interesting to
consider this solution in the space coordinates. The diagonal distribution
in momentum necessarily implies that the density matrix does not reduce to
a diagonal form with respect to space coordinates. The Fourier transform in
$Q$ and $q$ of $\rho_{d}(Q,q,t)$ gives the density matrix in position
representation:
\begin{eqnarray}
\rho_{d}(R,r,t)&=&2\sqrt{\pi\over M(\tau)}\exp\biggl[-\Bigl[{1\over 4\sigma^2}
e^{-2\tau}+{D\over 8\hbar^{2}\gamma}\nonumber\\
&&(1-e^{-2\tau})\Bigr]r^{2}
+\Big[ i\overline pe^{-\tau}\mp{i\epsilon\over\hbar\gamma}(1-e^{-2\tau})
\Bigr]r \nonumber\\
&-&{1\over M(\tau)}\Bigl[R-{\overline p\hbar\over m\gamma}(1-e^{-2\tau})
\nonumber\\
&&\pm{\epsilon\over m\gamma^2}(1-e^{-2\tau}-\tau)
-{i\hbar r\over 2\sigma^{2}m\gamma}e^{-2\tau}\nonumber\\
&&(1-e^{-2\tau}) +{iDr\over 4m\gamma^{2}\hbar}(1-e^{-2\tau})^{2}\biggr],
\label{18}
\end{eqnarray}
where $\tau = \gamma t$ and
\begin{eqnarray}
M(\tau) &=& \sigma^{2}+{\hbar^{2}\over\sigma^{2}m^{2}\gamma^2}
(1-e^{-\tau})^{2}\nonumber\\
&&+{D\over 2m^{2}\gamma^{3}}(2\tau-3+4e^{-\tau}-e^{-2\tau}).\label{19}
\end{eqnarray}
As $t\rightarrow\infty$, one can see that the off-diagonal elements of the
density matrix do not vanish, and the diagonal elements give the position
distribution function obtained by setting $r = 0$ and $R = x$:
\begin{eqnarray}
|\psi(x)|^{2} &=& 2\sqrt{\pi\over M(\tau)}\exp\left({-1\over M(\tau)}
(x-{\overline p\hbar\over m\gamma}\pm{\epsilon(1-\tau)\over m\gamma^2}
\right)^{2}.\nonumber\\
&&\label{20}
\end{eqnarray}
The results corresponding to $\pm\epsilon$ are for the up and down spins.
The centers of the position distribution function shift with time and are
clearly different for up and down spins. Though the distribution
$\rho_{d}(R,r,t)$ implies a nonlocality through its dependence on $r$, we
observe that the width $w_{1}$ of the distribution in $r$ is considerably
smaller than the width $w_{2}$ of the distribution in $R$. In the large
$\tau$ limit, $w_{2}/w_{1} =\hbar^{2} m^{2}\gamma^{4} /D^{2}$. If
$\rho_{d}(R,r,t)$ is coarse-grained over length scales $l$, such that $l$
is larger than the deBroglie wavelength of the particle $\gamma/\epsilon$
and $w_{1}$, but smaller than $w_{2}$ , we should have a local distribution
in position space. This means $l > Max(\gamma/\epsilon, \gamma/D)$, and $l
< D\hbar^{2}\tau/m^{2}\gamma^{2}$, which is surely possible for large enough
 $\tau$.

Thus if the initial wavefunction of the system-apparatus is a
product of the Gaussian wave-packet of (\ref{8}) and the apparatus state
$(a|\uparrow \rangle + b |\downarrow\rangle )$, {\it i.e.},
\begin{equation}
\psi(x,0) = {1\over(\sigma\sqrt{\pi})^{1/2}}[a|\uparrow\rangle +
b|\downarrow\rangle ] \exp(-i\overline px-x^{2}/2\sigma^{2}),\label{21}
\end{equation}
in the model without the environment, the time evolution of the density
matrix is
\begin{eqnarray}
\rho &=& |a|^{2}|\uparrow\rangle\langle\uparrow |\psi^{*}_{+}(x,t)\psi_{+}(y,t)
\nonumber\\
&&+ |b|^{2}|\downarrow\rangle\langle\downarrow
|\psi^{*}_{-}(x,t)\psi_{-}(y,t) \nonumber\\
&&+ ab^{*}|\uparrow\rangle\langle\downarrow |\psi^{*}_{+}(x,t)\psi_{-}(y,t)
\nonumber\\
&&+ a^{*}b|\downarrow\rangle\langle\uparrow |\psi^{*}_{-}(x,t)\psi_{+}(y,t),
\label{22}
\end{eqnarray}
where $\psi_{\pm}(x,t)$ are the wavefunctions of the particle in the
potential $\pm\epsilon x$. The environment causes the decay of the
off-diagonal elements and the large time limit of the density matrix
assumes the form
\begin{equation}
\rho_{R} = |a|^{2}|\uparrow\rangle\langle\uparrow |\rho_{\uparrow\uparrow}
+ |b|^{2}|\downarrow\rangle\langle\downarrow |\rho_{\downarrow\downarrow},
\label{23}
\end{equation}

with $\rho_{\uparrow\uparrow}$ and $\rho_{\downarrow\downarrow}$ being
given by (\ref{17}) in the momentum representation, and by (\ref{18}) in
the coordinate representation. This calculation clearly establishes the
measurement of spin via a momentum measurement. The spin diagonal density
matrix evolves to a diagonal form in the momentum space, while the spin
off-diagonal density matrix goes to zero with time. Further, the
probability distributions of up and down spins are also given by the
initial amlitudes $a$ and $b$ according to the quantum prescription. Figs.
1(a), (b) and (c) show the real part of the sum of the density matrices,
$\rho = Real(\rho_{\uparrow\uparrow} + \rho_{\downarrow\downarrow})$, given
by (\ref{17}) in the momentum representation for the diagonal spin elements
for three different values of the scaled time $\tau=\gamma t$. As $\tau$
increases, the off-diagonal elements clearly decay, leaving a diagonal
distribution which is centered around two different mean momenta,
corresponding to up and down spins. One can identify these mean momenta as
those with which the centers of the wave packets corresponding to up and
down spins move.

\section{Summary of results}
In this investigation we have considered through a canonical example two
essential aspects of the measurement problem, which are (a) decoherence of
the superpositions in the apparatus states so as to allow classical
inference, and (b) the definite correlation of the system states with the
apparatus states.  We find that measurement is achieved if the apparatus is
macroscopic enough to be affected by an environment and furthermore, its
relevant degree of freedom has a classical limit in the sense of the
Correspondence principle. This is to be contrasted with the case in which
the relevant degree of freedom of the apparatus has a discrete spectrum,
because in that situation the correlation between the system and the
apparatus states is not achieved.  To conclude, we have been able to provide
a scheme of incorporating a concept like "classical apparatus" in a purely
quantum formalism and demonstrate that a suitable quantum apparatus when
dissipatively coupled to an appropriate environment does perform a
measurement.  It is in this sense that we justify the concept of Bohr and
von Neumann that a measurement requires the interaction of a quantum system
with a classical system.

\begin{acknowledgements}
DK acknowledges several illuminating discussions with Dr. Ishwar Singh.
AV acknowledges financial support from the University Grants Commission,
India.
\end{acknowledgements}

\appendix
\section{}
Equation (\ref{6}) is equivalent to the following set of ordinary differnetial
equations :
\begin{eqnarray}
{dt\over ds} &=& 1,\label{A1}\\
{dr\over ds} &=& \gamma (r-r_{Q}),\label{A2}\\
{d\rho_{d}\over ds} &=& -\rho_{d}\left( {Dr^{2}\over 4\hbar^{2}} \mp
{i\epsilon r\over\hbar}\right),\label{A3}
\end{eqnarray}
with $r_{Q}=\hbar Q/m\gamma$. The invariants of these orbits with respect
to $s$ are easily found to be
\begin{equation}
I_{1} = (r-r_{Q}) e^{-\gamma t},\label{A4}
\end{equation}
and
\begin{eqnarray}
I_{2} &=& \rho_{d}\exp\biggl[{D\over 4\hbar^{2}}\left( r^{2}_{Q}t
+ {2r_{Q}\over\gamma}(r-r_{Q}) +{(r-r_{Q})^{2}\over
2\gamma}\right)\nonumber\\
&& \pm {i\epsilon\over\hbar}\left( r_{Q}t + {{r-r_{Q}}\over\gamma}\right)
\biggr].\label{A5}
\end{eqnarray}
Clearly, $I_{2} = w(I_{1})$, where $w$ is an arbitrary function. This
enables us to write
\begin{eqnarray}
\rho_{d}(Q,r,t) &=& w(I_{1})\exp\biggl[{D\over 4\hbar^{2}}
\Bigl( r^{2}_{Q}t + {2r_{Q}\over\gamma}(r-r_{Q})\nonumber\\
&& +{(r-r_{Q})^{2}\over
2\gamma}\Bigr) \pm {i\epsilon\over\hbar}\left( r_{Q}t +
{{r-r_{Q}}\over\gamma}\right) \biggr],\label{A6}
\end{eqnarray}
$w(I_{1})$ is now determined from the initial condition (\ref{8}) for
$\rho_{d}(Q,r,0)$. One can easily see that
\begin{eqnarray}
w(I_{1}) &=& w\left( (r-r_{Q})e^{-\gamma t}\right) \nonumber\\
&=& \exp\biggl[i\overline p(r-r_{Q})e^{-\gamma t} \nonumber\\
&&- {1\over 4\sigma^{2}}
\{(r-r_{Q})^{2}e^{-2\gamma t} + 2r_{Q}(r-r_{Q})e^{-\gamma t}\}\nonumber\\
&& + {D\over 4\hbar^{2}}\Bigl({2r_{Q}\over\gamma}(r-r_{Q})e^{-\gamma
t}\nonumber\\
&&+ {(r-r_{Q})^{2}e^{-2\gamma t}\over 2\gamma}\Bigr) \pm {i\epsilon\over\hbar}
(r-r_{Q})e^{-\gamma t}\biggr].\label{A7}
\end{eqnarray}
Substituting this in (\ref{A6}) gives the result (\ref{12}) for
$\rho_{d}(Q,r,t)$. To
solve (\ref{7}) we first make the transformation
\begin{equation}
\rho_{od} = W\exp(\mp 2i\lambda t).\label{A8}
\end{equation}
The equation for $W(Q,r,t)$ is now equivalent to the following set of
differential equations :
\begin{eqnarray}
{dt\over ds} &=& 1, \label{A9}\\
{dr\over ds} &=& \gamma (r-r_{Q}), \label{A10}\\
{dQ\over ds} &=& \pm 2\epsilon /\hbar, \label{A11}\\
{dW\over ds} &=& {-Dr^{2}\over4\hbar^{2}}W.\label{A12}
\end{eqnarray}
The invariants for this set of equations are :
\begin{eqnarray}
I_{1} &=& \hbar /m\gamma(Q\pm 2\epsilon t/\hbar) \label{A13}\\
I_{2} &=& (r-\hbar Q/\gamma m\pm 2\epsilon /m\gamma^{2})e^{-\gamma t}.
\label{A14}
\end{eqnarray}
The third invariant from (\ref{A11}) is obviously a finction of $I_{1}$ and
$I_{2}$, hence,
\begin{eqnarray}
W(Q,r,t)&=& f(I_{1},I_{2})\exp\biggl[-{D\over 4\hbar^{2}}\Bigl\{ \left(I_{1}
\mp{2\epsilon\over m\gamma^{2}}\right)^{2}t \nonumber\\
&& + 2\left(I_{1}\mp{2\epsilon\over m\gamma^{2}}\right)\left(
{I_{2}e^{-\gamma t}\over\gamma}\mp{\epsilon t^{2}\over m\gamma}\right)
+ {I_{2}^{2}e^{-2\gamma t}\over 2\gamma}\nonumber\\
&&\mp {4I_{2}\epsilon\over m\gamma^{3}}
e^{-\gamma t}(\gamma t -1)+{4\epsilon^{2}t^{3}\over 3m^{2}\gamma^{2}}\Bigr\}
\biggr]. \label{A15}
\end{eqnarray}
$f(I_{1},I_{2})$ can now be easily determined from the initial condition
(\ref{8})
for $\rho(Q,r,0)$:
\begin{eqnarray}
f(I_{1},I_{2})&=&\exp\biggl[-{\sigma^{2}\over 4}(Q\pm{2\epsilon t\over\hbar}
)^{2} - {1\over 4\sigma^{2}}\Bigl\{ \left({\hbar Q\over m\gamma} \mp
{2\epsilon t\over\gamma^{2}}\right)\nonumber\\
&&(1-e^{-\gamma t}) + re^{-\gamma t} \pm {2\epsilon t\over\gamma}\Bigr\}^{2}
+ i\overline p\Bigl\{ \Bigl({\hbar Q\over m\gamma}\nonumber\\
&&\mp {2\epsilon t\over\gamma^{2}}\Bigr)
(1-e^{-\gamma t})+ re^{-\gamma t} \pm {2\epsilon t\over\gamma}\Bigr\}
\biggr] \nonumber\\
&&\exp\biggl[ {D\over 2\hbar^{2}\gamma}\left({\hbar Q\over m\gamma}
\pm {2\epsilon t\over\gamma^{2}} \mp {2\epsilon\over m\gamma^{2}}\right)
\Bigl( r-{\hbar Q\over m\gamma}\nonumber\\
&&\pm {2\epsilon\over m\gamma^{2}}\Bigr)
e^{-\gamma t}+{D\over 2\hbar^{2}\gamma}\Bigl( r-{\hbar Q\over m\gamma}
\pm {2\epsilon\over m\gamma^{2}}\Bigr)^{2}e^{-\gamma t}\nonumber\\
&&\pm {D\epsilon\over m\gamma^{3}}\Bigl( r-{\hbar Q\over m\gamma}
\pm {\epsilon\over m\gamma^{2}}\Bigr)e^{-\gamma t}\biggr].\label{A16}
\end{eqnarray}
Substituting for $f(I_{1},I_{2})$ in (\ref{A15}) gives the result of (\ref{9}).

\begin{figure}
\caption{Plot of the sum of the real part of the spin-diagonal density
matrices in the momentum representation, $\rho = Real(\rho_{\uparrow\uparrow }
+ \rho_{\downarrow\downarrow })/\sigma$, given by Eq.(11), versus dimensionless
momenta $u = (q+Q/2)\sigma$ and $v = (q-Q/2)\sigma$, for (a) $\tau\equiv
\gamma t = 0$, $\epsilon /m\gamma^{2}=0.0$, (b) $\tau = 1$, $\epsilon
/m\gamma^{2}=2.0$, (c) $\tau = 3$, $\epsilon /m\gamma^{2}=2.0$, with
$\overline p=0.2/\sigma$, $D/m^{2}\gamma^{3}=\sigma^{2}$, $m\gamma
/\hbar=0.5/\sigma^{2}$.}
\end{figure}

\epsfysize=2.4in
\centerline{\epsfbox{n1.ps}}
\centerline{Figure 1(a)}
\bigskip

\epsfysize=2.4in
\centerline{\epsfbox{n2.ps}}
\centerline{Figure 1(b)}
\bigskip

\epsfysize=2.4in
\centerline{\epsfbox{n3.ps}}
\centerline{Figure 1(c)}
\bigskip

\end{narrowtext}
\end{document}